\documentstyle[prl,aps,epsfig]{revtex}
 \newcommand{\zr}[1]{\mbox{\hspace*{#1em}}}
 \newcommand{\ID}{\mbox{{\sf 1}\zr{-0.14}\rule{0.04em}{1.55ex}\zr{0.1}}}

 \newcommand{\ZZ}{\mbox{\sf Z\zr{-0.45}Z}}

\newcommand{\gl}{\mathrel{\rlap{\lower2pt\hbox{\small \hskip1pt$<$}}
\hspace*{1pt}\raise3pt\hbox{\small $>$}}}

\begin{document} 
\draft
\twocolumn[\hsize\textwidth\columnwidth\hsize\csname@twocolumnfalse%
\endcsname
\preprint{hep-lat/0107018}

\title{'t~Hooft Loops, Electric Flux Sectors and Confinement 
in $SU(2)$ Yang-Mills Theory} 

\author{Philippe de Forcrand}
\address{Institut f\"ur Theoretische Physik, ETH H\"onggerberg,
  CH-8093 Z\"urich, Switzerland}
\address{CERN, Theory Division, CH-1211 Gen\`eve 23, Switzerland}
\author{Lorenz von Smekal}
\address{Institut f\"ur Theoretische Physik III, Universit\"at
  Erlangen-N\"urnberg, Staudtstr. 7, D-91058 Erlangen, Germany} 
\date{\today}
\maketitle
\begin{abstract}
We use 't~Hooft loops of maximal size on finite lattices to
calculate the free energy in the sectors of $SU(2)$ Yang-Mills theory 
with fixed electric flux as a function of temperature and
(spatial) volume. Our results provide evidence for
the mass gap. The confinement of electric fluxes 
in the low temperature phase and their condensation in the high temperature
phase are demonstrated. In a surprisingly large
scaling window around criticality, the transition is quantitatively well
described by universal exponents and amplitude ratios relating the properties
of the two phases. 
\end{abstract}
\pacs{PACS numbers:   12.38.Gc, 12.38.Aw, 11.15.Ha 
              \hfill  CERN-TH/2001-197,  FAU-TP3-01/7}
]

Center symmetry is widely believed to play a key role for confinement.
In $SU(N)$ Yang-Mills theory at finite temperature, the Polyakov loop 
is commonly used in lattice studies to illustrate this role.
Its correlations are short range at low temperature and 
acquire a non-zero disconnected part above
a critical temperature $T_c$, signaling deconfinement.
The role of the low (high) temperature phase as the center symmetric (broken)
one is thereby the opposite from that in the corresponding $\ZZ_N$-spin
model. This suggests to consider dual variables, 
whose behavior as a function of temperature is reversed.
That point of view was emphasized by 't~Hooft with the
introduction of the dual to the Wilson 
loop \cite{tHo78}. Temporal 't~Hooft loops in $SU(2)$ show 
screening for the interaction of a static
pair of center monopoles in both phases, for $T<T_c$ and for $T >
T_c$ \cite{Kov00,deF00}, just as spatial Wilson loops exhibit an area
law in either case. The expectation values  of (sufficiently large) 
spatial 't~Hooft loops  $\widetilde W(C)$, on the other hand, change from 
a screening behavior below $T_c$ to a confined one \cite{Kor99} with a
dual string tension $\tilde\sigma$ and an area law 
in the ``electrically'' deconfined phase above $T_c$ \cite{deF00}, 
\begin{equation}
  \label{eq:1}
   \langle \widetilde W(C) \rangle \, \sim \, \exp\{  -
   \tilde\sigma(T) \, L R  \} \; , \quad \mbox{at $T>T_c$}\, , 
\end{equation}
for a rectangular curve $C$ spanning a spatial surface of size $L
\times R$ on a $1/T \times L^3$ lattice. Ref.~\cite{deF00} also
confirmed numerically a perimeter law for 't~Hooft loops at $T=0$. 

The qualitative behavior of the spatial 't~Hooft loops 
is thus the same as that of the Wilson loops in the 3-dimensional 
$\ZZ_2$ gauge theory  \cite{Sav80}. 
Furthermore, as the phase transition is approached (from above),
the temperature dependence of the $SU(2)$ dual string tension
obeys the same scaling law as that of the interface tension in the
3-dimensional Ising model (below $T_c$). This similarity is expected, since
the 3d $\ZZ_2$ gauge theory, its  dual the 3-dimensional Ising model,
and $SU(2)$ at finite temperature all belong to the same universality class.  

To introduce a spatial 't~Hooft loop 
of maximal size $L\times L $, living in, say, the $(x,y)$
plane of the dual lattice, one multiplies one plaquette in every $(z,t)$ 
plane of the original lattice by a non-trivial element of the center 
of $SU(N)$ in such a way that the modified plaquettes form a coclosed set. 
This creates a $\ZZ_N$-interface which is equivalent to
enforcing boundary conditions with twist in the $(z,t)$ directions. 
Combining two and three maximal $L\times L$ 't~Hooft loops in orthogonal
spatial planes yields the partition functions of $SU(N)$ Yang-Mills theory 
on finite lattices for all possible combinations of temporal twists.
From these, one obtains the free energies in the presence of fixed
units of electric flux $\vec e\in \ZZ_N^3$ via
a $\ZZ_N$ Fourier transform as shown by 't~Hooft \cite{tHo79}.
This leads to the respective electric-flux superselection
sectors of the theory in the thermodynamic limit. 
In the present paper, we verify for $SU(2)$ that the partition functions of
electric fluxes $\vec e  \not=0$ vanish exponentially with the spatial 
size $L$ of the system for $T<T_c$ in the magnetic Higgs phase with electric
confinement, whereas for $T>T_c$ they become equal to that of the neutral 
$\vec e = 0$ sector. Describing the transition by critical
exponents of the 3d-Ising class, we discuss how universal
amplitude ratios quantitatively relate the two phases, in particular,
the string tension below and the dual (vortex) string
tension above $T_c$.

As expected, the free energy $F_q$ of a 
static fundamental charge jumps from $+\infty$ to $0$ at $T_c$.
One might expect to see this also in measuring 
the Polyakov loop $P$ directly on the lattice. 
If $\langle P \rangle \equiv e^{-\frac{1}{T} F_q}$, 
an infinite free energy amounts to a center symmetric distribution, while
a non-zero expectation value $\langle P \rangle$ is obtained
for finite $F_q$. 
However, the presence of a single charge is incompatible with
periodic boundary conditions to measure $\langle P \rangle$. 
And, like any Wilson loop, $\langle P \rangle$ is subject to UV-divergent
perimeter terms, such that $\langle P \rangle = 0$ at all $T$
as the lattice spacing $a \!\rightarrow\! 0$.
Here, we measure the gauge-invariant, UV-regular
free energy of a static fundamental charge in $SU(2)$, and show that it has
the expected behaviour, dual to that of a certain type of center vortex.
Both provide a well-defined order parameter for the transition \cite{deF01}.

\noindent{\it Twisted Boundary Conditions and Electric Fluxes}  
\medskip

For the finite-volume partition functions of the pure $SU(N)$ gauge theory,
't~Hooft's twisted boundary conditions fix the total number of
$\ZZ_N$-vortices modulo $N$ that pierce planes of a given orientation. 
Thus, on the 4-dimensional torus there are $N^6$ different $\ZZ_N$-flux 
sectors corresponding to the 6 possible orientations for the planes of the
twists.
These label the inequivalent choices for imposing boundary conditions
on the gauge potentials $A$, which are invariant under the center $\ZZ_N$ of
$SU(N)$.
One first chooses $A(x)$ to be periodic with the lengths $L_\mu$ 
of the system in each direction $\mu$ 
up to gauge transformations $\Omega_\mu(x_\perp) \in SU(N)$ which
can depend on the components $x_\perp$ transverse to that direction,
\begin{equation} 
 A(x\!+\!L_\mu) =  \,
          \Omega_\mu(x_\perp) \, \big(A(x) - \frac{i}{g} \, \partial \big)\,  
        \Omega_\mu^{-1}(x_\perp)   
     \; .       \label{tbcs}
\end{equation} 
Then, compatibility of two successive translations in a ($\mu,\nu$)-plane
entails that (no summation of indices) 
\begin{eqnarray}
&&{\Omega_\mu(x_\perp\! +\! L_\nu) \Omega_\nu(x_\perp) = Z_{\mu\nu} \,
 \Omega_\nu(x_\perp\! +\!  L_\mu)       \Omega_\mu(x_\perp)}
\nonumber\\
   &&{  \mbox{with}  \;\; Z_{\mu\nu} = e^{2\pi i n_{\mu\nu}/N} \;, 
   \mbox{ and }  n_{\mu\nu} = - n_{\nu\mu}  \in \ZZ_N } \; .  \label{coc}
\end{eqnarray} 
The total number mod.~$N$ of center vortices in a ($\mu,\nu$)-plane
is specified  in each sector by the corresponding component
of the twist tensor $n_{\mu\nu}$. 
The spatial ones are given by the conserved, $\ZZ_N$-valued and gauge-invariant
magnetic flux $\vec m$ through the box, $n_{ij} \equiv \epsilon_{ijk} m_k $.
The time components $ n_{0i} \equiv  k_i$ define {temporal twist} 
$\vec k \in \ZZ_N^3$.

With the inequivalent choices of boundary conditions, 
the finite-volume theory decomposes into sectors of fractional 
Chern-Simons number ($\nu + \vec k\cdot\vec m /N$) \cite{vBa82} 
and states labelled by $|\vec k , \vec m , \nu \rangle $, where $\nu
\in \ZZ$ is the usual instanton winding number.
However, these sectors are not invariant under homotopically non-trivial
gauge transformations $\Omega [\vec k,\nu]$ which can change $\vec k$ and
$\nu$,  
\begin{equation}
\label{hom-nt}
      \Omega [\vec k',\nu'] \, |\vec k , \vec m , \nu \rangle \, =\,  
          |\vec k\! +\!\vec k', \vec m , \nu\! +\!\nu' \rangle \, .
\end{equation}
A Fourier transform of the twist sectors $Z(\vec k, \vec m, \nu)$ which
generalizes the construction of $\theta$-vacua as 
Bloch waves from $\nu$-vacua in two ways, 
by replacing $\nu \to (\nu + \vec k\cdot\vec m /N)$ 
for fractional winding numbers and with an additional
$\ZZ_N$-Fourier transform w.r.t.~the temporal twist $\vec k$, 
\begin{equation}
 e^{-\frac{1}{T} F(\vec e, \vec m,\theta)}  \,=\, \frac{1}{N^3} \, \sum_{\vec
 k, \,\nu} \,  e^{-i\omega(\vec k, \nu)} \,
 Z(\vec k,  \vec m, \nu)  \, ,  \label{ZnFT}
\end{equation}    
yields the free energy $ F(\vec e, \vec m,\theta)$ in an ensemble of
states invariant, up to a geometric phase $ \omega(\vec k,\nu) = 2\pi \vec
e\cdot\vec k/N  + \theta (\nu \!+\! \vec k\cdot\vec m /N)$,  
under the non-trivial $\Omega[\vec k,\nu]$ also,
\begin{equation}
\label{hom-nt-inv}
      \Omega [\vec k,\nu] \, |\vec e , \vec m , \theta \rangle \, =\,  
         \exp\{ i \omega(\vec k,\nu) \} \, 
  |\vec e , \vec m , \theta \rangle \, .
\end{equation}
These states are then classified, in addition to their magnetic flux $\vec m $
and vacuum angle $\theta $, by their $\ZZ_N$-valued  
{\em gauge-invariant electric flux} in the $\vec e$-direction \cite{tHo79}.

\begin{figure}[t]
\vspace*{-.5cm}
  \centering{\
        \epsfig{file=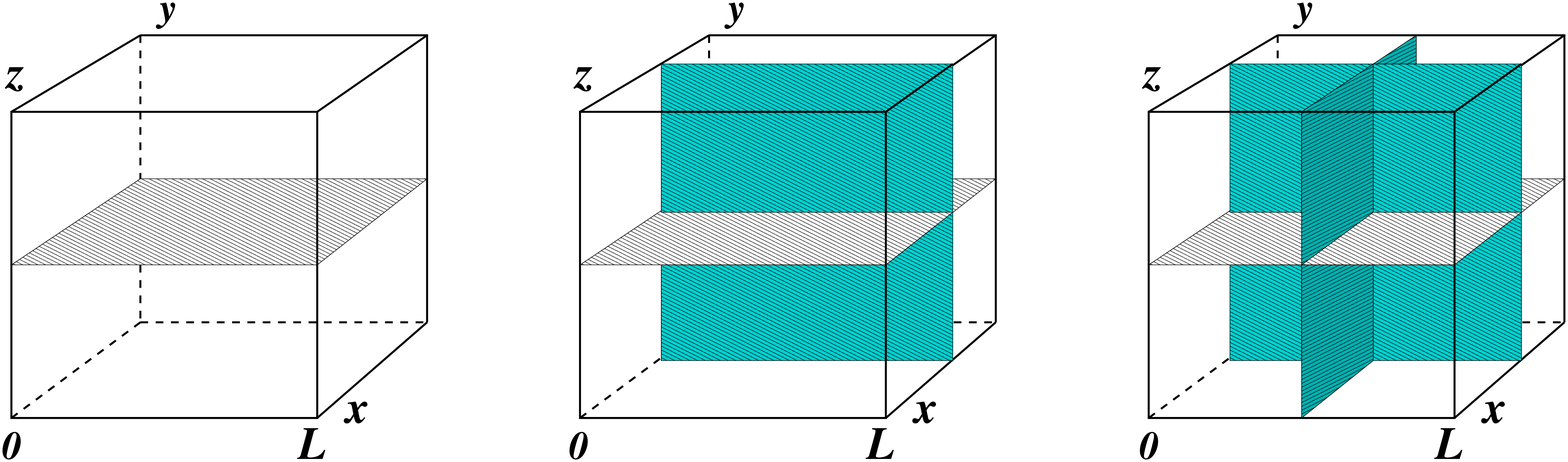,width=\linewidth}}
\vspace{.2cm}
\caption{Cubes with one, two, and three $L\times L$ planes
dual to the stacks of flipped plaquettes, sketched for temporal 
twist $\vec k = (0,0,1)$, $\vec k = (0,1,1)$, and
$\vec k = (1,1,1)$ from left to right.}  
\label{tHloops} 
\end{figure}

To create $n_{\mu\nu}$-twist in $SU(2)$, on the lattice, 
one introduces couplings with reversed signs
into the usual Wilson action by replacing $\beta \to -\beta $ for one
plaquette in every $(\mu ,\nu)$-plane,
\begin{equation}
  S(\beta,\vec k,\vec m) = 
       - \sum_P  \beta (P) \,  \frac{1}{2} \, \mbox{Tr}(U_P)  \; .
\end{equation}
Herein, the sum extends over all plaquettes $P$  with $U_P$ denoting 
the path-ordered product of the links around $P$ and 
we introduced a plaquette-dependent coupling,
\begin{equation}
              \beta (P)  = \left\{ { 
                            - \beta \; , \;\; P \in {\cal P}(n_{\mu\nu})
                            \atop 
                            \phantom{-}  \beta \; , \;\; P \not\in
                            {\cal P}(n_{\mu\nu}) 
                            } \right.
\end{equation} 
where ${\cal P}(n_{\mu\nu})$ denotes the coclosed stacks of plaquettes dual to 
the planes of the maximal 't~Hooft loops. 
These stacks of flipped plaquettes force the $\ZZ_2 $-interfaces 
corresponding to twist in the $(\mu ,\nu)$-directions.
Equivalently, they create a 't~Hooft loop of maximal size in the 
orthogonal plane.
For the various combinations of temporal twist the coclosed sets ${\cal
P}(\vec k)$, when put between two time-slices, can be chosen dual to the
spatial planes shown in Fig.~\ref{tHloops}. 

The partition functions of the twist sectors, relative to 
the untwisted $Z_\beta $ (such that $Z_\beta(\vec 0,\vec 0)  = 1$), are then:
\begin{equation}  
Z_\beta(\vec k,\vec m) = Z_\beta^{-1} \int [dU] \,  \exp- S(\beta,\vec k,\vec
m)\; . \label{Zb}
\end{equation}
Similarly, flipping the couplings of plaquettes dual to some surface subtended
by a closed curve $C$ yields the expectation value of a finite 't~Hooft
loop $\widetilde W(C)$. 
An entirely analogous procedure can be used for
Wilson loops in the 3-dimensional $\ZZ_2$ gauge theory. 
Through duality, their expectation values can be expressed as ratios of
Ising-model partition functions with and without antiferromagnetic bonds at
those links of the Ising-model that are dual (in 3 dimensions) to some
surface spanned by the $\ZZ_2$-Wilson loop \cite{Sav80}. 
In both cases the surface is arbitrary except for its
boundary. 
The 't~Hooft loop can thus be viewed as a gauge-invariant operator which
creates a fluctuating center-vortex surface with pinned boundary.

In this paper, we first calculate $Z_k(\vec k) \equiv Z_\beta (\vec k,0) $
for $\vec m  =0$ (and  $\theta =0$) on a $1/T \times L^3 $ lattice in $SU(2)$
(with $k_i$ in $\{0,1\}$). The expectation values of maximal-size 't~Hooft
loops, given by the partition functions of the twist sectors, are then 
used to calculate the free energies of electric fluxes as per
Eq.~(\ref{ZnFT}). For purely temporal twists in particular, 
the free energies of the electric fluxes through the $L^3$ box at temperature
$T$, $ F_e(\vec e;L,T) \equiv F(\vec e,\vec m\!=\!0,\theta\!=\!0) - F(\vec
e\!=\!0,\vec m\!=\!0,\theta\!=\!0)$, are given by: 
\begin{equation}
\label{ZeFT}
       Z_e(\vec e) \equiv  e^{-\frac{1}{T} F_e(\vec e;L,T)} = 
\frac{\sum_{\{k_i = 0\}}^{N-1} \,  e^{-2\pi i \, \vec e \cdot \vec k
 /N } \, Z_k(\vec k)}{\sum_{\{k_i = 0\}}^{N-1} \, Z_k(\vec k)}  
\end{equation}
with $Z_e(\vec 0) = Z_k(\vec 0) = 1$. 
Because of the invariance under spatial $\pi/2$ rotations,
we can write for $SU(2)$:
$Z_k(1)$ if $\vec k = \{(1,0,0),(0,1,0),(0,0,1)\}$; $Z_k(2)$ if 
$\vec k = \{(1,1,0),$ $(1,0,1),(0,1,1)\}$; and $Z_k(3) = Z_k(1,1,1)$,
for the partition functions with one, two and three maximal
't~Hooft loops in orthogonal spatial planes, respectively. With analogous
notations for the $Z_e(\vec e)$ one thus obtains: 
\begin{eqnarray}
     Z_e(1) &=& \frac{1}{{\cal N}_e} \Big( 1 + Z_k(1)
           -   Z_k(2) -  Z_k(3) \Big)   \; , \label{Ze1} \\
     Z_e(2) &=& \frac{1}{{\cal N}_e} \Big( 1 - Z_k(1)
           -   Z_k(2) + Z_k(3) \Big)   \; , \label{Ze2} \\
     Z_e(3) &=& \frac{1}{{\cal N}_e} \Big( 1 - 3 Z_k(1)
           + 3  Z_k(2) -  Z_k(3) \Big)   \; , \label{Ze3} \\[2pt]
    \mbox{with } && {\cal N}_e \,=\,   1 + 3 Z_k(1)
           + 3 Z_k(2) + Z_k(3) \; . \nonumber
\end{eqnarray}      
Eqs.(\ref{Ze1})-(\ref{Ze3}) are readily inverted via inverse $\ZZ_2$ Fourier
transform, which in effect interchanges $ Z_e(i) \leftrightarrow Z_k(i)$.

We now establish the connection with Polyakov loops.
First, recall that the gauge invariant definition of the latter in the
presence of temporal twists is given by 
\cite{vBa84},
\begin{equation}
\label{Pdef}
      P(\vec x) =  \frac{1}{N} \mbox{tr} \, \Big( {\mathcal{P}} e^{ig
      \int_0^{1/T}  A_0(\vec x,t) dt } \, \Omega_t(\vec x) \Big) \; .
\end{equation}
Successively transforming the path-ordered exponential by the various spatial
transition functions which accompany the possibly non-trivial
$\Omega_t(\vec x)$ for the transition in the time direction, we 
derive from (\ref{tbcs}) and (\ref{coc}) that   
\begin{eqnarray}
\label{Pbc}
     P(\vec x + L \vec e) \, = \, e^{2 \pi i \, \vec e \cdot \vec k/N} \,
     P(\vec x)  \;, \; \; \mbox{or} \nonumber \\  
      P(\vec x)  P^\dagger(\vec x + L \vec e)  \,
     = \, e^{-2 \pi i \, \vec e \cdot \vec k/N} \, \ID  \; .
\end{eqnarray}
This is proportional to the unit operator when acting on the states  
in a sector of definite $\vec k$-twist.
Therefore, the electric-flux partition functions of Eq.~(\ref{ZeFT}) 
are in fact the expectation values of Polyakov loop correlators 
in the ensemble average over all these temporal twists,
\begin{equation}
\label{ZeP}
  Z_e(\vec e) =  e^{-\frac{1}{T} F_e(\vec e;L,T)} = 
        \big\langle P(\vec x) P^\dagger(\vec x +L\vec e) \big\rangle _{L,T}
        \; .
\end{equation}
This expectation value is taken in the no-flux ensemble,
with enlarged partition function $Z = \sum_{\{k_i = 0\}}^{N-1}
\,  Z_k(\vec k) $,
which is manifestly different, in a finite volume, from the periodic
ensemble. Also note that the operator in Eq.~(\ref{ZeP}) has no perimeter, 
is UV-regular, and we will see that there is no Coulomb term for small
volumes either.

\begin{figure}[t]
\vspace*{-.7cm}
\rightline{\epsfig{file=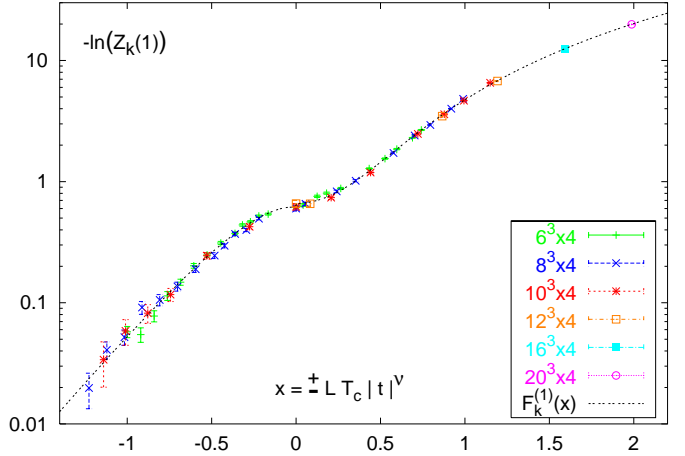,width=1.1\linewidth}}
\vskip -.5cm
\caption{The free energy of one temporal twist 
as a function of the finite size scaling variable $x$ (with $x<0$ 
for $T<T_c$).}   \label{tH1all} 
\end{figure}

Eq.~(\ref{ZeFT}) thus yields a dual relation between Polyakov loops and 
temporal twists of the general pattern,
\[
    \big\langle P(\vec x) P^\dagger(\vec x +L\vec e) \big\rangle
        \stackrel{L\to\infty}{\longrightarrow}  \left\{ { 
                  0 \, , \;\mbox{for} \; Z_k(\vec k) \to 1 \, , \; T < T_c
                                                   \atop 
                  1 \, , \;\mbox{for} \; Z_k(\vec k) \to 0 \, , \; T > T_c
                            } \right. 
\] 
reflecting the different realizations of the electric $\ZZ_N$ center symmetry
in the respective phases.

\medskip
\noindent{\it Finite Size Scaling and Universal Amplitude Ratios} 
\medskip

We compute the three partition functions $Z_k(i)$ for $SU(2)$ near $T_c$
by Monte Carlo, using the method of Ref. \cite{deF00}, with 20k--50k
measurements per simulation. 
For pioneering related work at $T=0$, see Ref. \cite{Has90}.
The $Z_k(i)$ are the analogues of ratios of 3d-Ising model partition
functions with different boundary conditions. As for the latter \cite{Has93}, 
we assume their $L$, $T$ dependence to be governed by 
simple finite-size scaling laws,

\vspace{-.1cm}

\begin{eqnarray} 
    \hskip 1cm Z_k(i) \, =  \, f_\pm^{(i)}(x) \; , \; i = 1,.. 3 \;
    . \label{fss} 
\end{eqnarray}

\vspace{-.2cm}

\noindent
The $f_\pm^{(i)}$ are functions of the finite-size scaling variable 
\begin{equation}
 \hspace*{1.4cm} x = \pm L T_c |t|^\nu  \propto  L/\xi_{\pm}(t)      \; , \;
 \mbox{for} \;  T \gl  T_c \; , 
\end{equation}
where $ t = T/T_c - 1 $ and $\xi_{\pm}(t) = \xi^0_\pm  |t|^\nu$   
are the reduced \linebreak temperature and the correlation lengths,
respectively, and we use the exponent $\nu = 0.63$ from the Ising model.

In our calculations we keep the number of points $N_t$ 
in the time direction fixed and control the temperature by varying the
lattice coupling $\beta$ around the critical value for the phase transition
$\beta_c$. The results presented here are obtained with $N_t = 4$ for which 
$\beta_c = 2.29895(10)$ \cite{Eng96}. We employ 
couplings $\beta $ between $2.19$ and $2.5$ for various lattices 
ranging from $N_l = 6$ up to $N_l = 20$ points in the spatial directions. 
Our method to calculate a partition function for one 
temporal twist amounts to $N_l^2$ independent Monte Carlo simulations
\cite{deF00}. For a fixed statistical accuracy, which introduces another 
factor $N_l^2$, the cost of the calculation roughly increases as $ N_l^7$. 

For the temperature $T=1/(N_t a)$, 
where the lattice spacing $a\equiv a(\beta)$ depends on the coupling,
we adopt the leading scaling behavior around criticality of the form,
\begin{equation}
\label{Tdef}
                     T / T_c  \, =  \, \exp\{ b \, (\beta - \beta_c) \} \; .
\end{equation}
The non-perturbative coefficient herein, we use $b = 3.26$, is determined so
as to reproduce the published values of $\beta_c(N_t)$ also for $N_t=6$ and
$8$ \cite{Fin92}.

\begin{figure}[b]
\vspace*{-.5cm}
\rightline{\epsfig{file=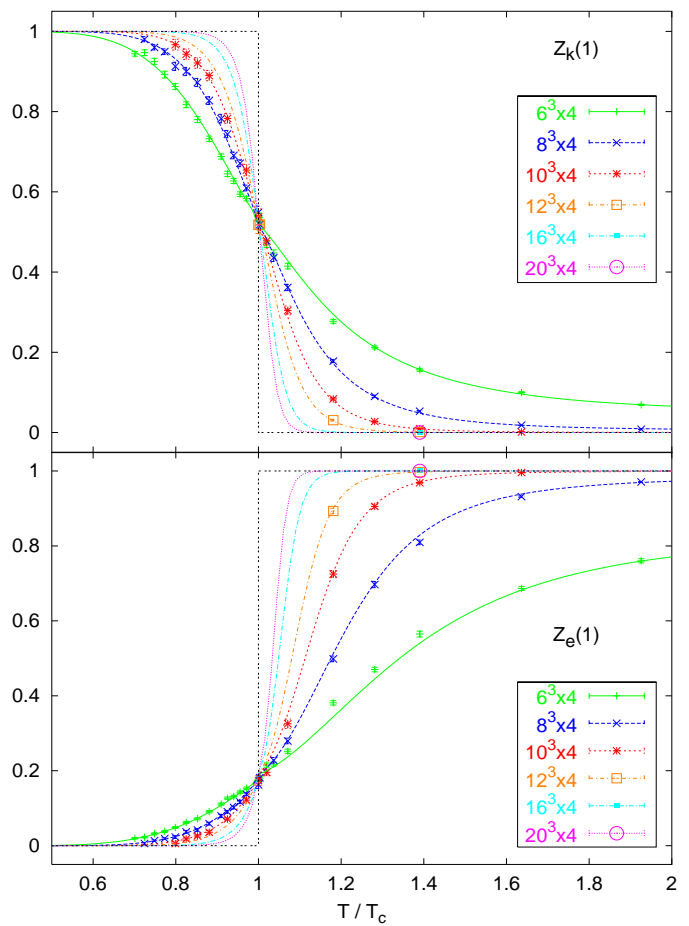,width=1.1\linewidth,height=13cm}}
\caption{The partition functions of one temporal twist (top) 
and one electric flux (bottom) over $T$ for the various lattices.}   
\label{ZkT} 
\end{figure}

We can see in  Fig.~\ref{ZkT} that all our curves intersect at the same point
$T\!=\!T_c$, which is consistent with the known value of $\beta_c $,
even for our smallest lattice size.  
In fact, this intersection point is known to provide a quite accurate
determination of the critical coupling  $\beta_c $ already on very small
lattices in the Ising model \cite{Has93}.

Away from criticality, corrections to (\ref{Tdef}) will become 
important if one is interested in a more precise definition of 
the physical temperature for $SU(2)$, as in Ref.~\cite{Fin92}. 
Especially at our lowest $\beta$ values, where such corrections are
noticable, scaling violations become important also.
A more refined definition of the reduced temperature $t$ is therefore beyond
the scope of our scaling analysis which concerns the dominant behavior near
$T_c$. It has no effect on the qualitative conclusions emphasized here.


That our results for all different lattice sizes nicely collapse on a single
curve can be seen for one spatial 't~Hooft loop in Fig.~\ref{tH1all}, 
with analogous results for $i\!=\!2$ and 3.
We fit the free energies by an Ansatz for each phase, where the 
two leading terms obey the expected thermodynamic behavior,
while the others represent an ad-hoc modeling of small-size corrections
(with $a_\pm^{\scriptscriptstyle (i)} \sim c_\pm^{\scriptscriptstyle (i)} < 
b_\pm^{\scriptscriptstyle (i)};\,  d_\pm^{\scriptscriptstyle (i)} < 1$),
\begin{eqnarray}
F_k^{(i)}(x) = \left\{ { \; \exp\Big\{b_-^{(i)} x + a_-^{(i)} -
    \frac{c_-^{(i)}}{(d_-^{(i)} - x)^2}\Big\}  \, , \;  x<0  \atop 
          b_+^{(i)} x^2 - a_+^{(i)} +
     \frac{c_+^{(i)}}{d_+^{(i)} + x^2} \, , \;\; x>0 } \right. &&
                  \label{f+-}
\vspace*{-.1cm}
\end{eqnarray} 
and $-\ln f_\pm^{(i)}(x)   \equiv F_k^{(i)}(x) $ for $x\gl 0$. 
Plotted over temperature, the data of Fig.~\ref{tH1all} and the unique function
$F_k^{\scriptscriptstyle (1)}(x)$ lead
to the family of curves shown in Fig.~\ref{ZkT} (top)
in which the phase transition is exhibited most clearly.
The amplitudes relevant to the large size behavior of the
free energy near criticality come out as
$b_-^{\scriptscriptstyle (1)} = 3.87 \pm 0.5$ and 
$b_+^{\scriptscriptstyle (1)} =  \tilde\sigma_0^{\scriptscriptstyle (1)} = 5.36
\pm 0.1$, with some additional systematic uncertainty inherent in the form of
our Ansatz (\ref{f+-}). With analogous data and fits for $Z_k(2)$ and
$Z_k(3)$,  from Eq.~(\ref{Ze1}), we obtain $Z_e(1)$ as 
shown in Fig.~\ref{ZkT} (bottom). Corrections to scaling 
do not become appreciable up to $T\sim 2T_c$ 
indicating a surprisingly large scaling window.

\begin{figure}[t]
\vspace*{-.7cm}
\rightline{\epsfig{file=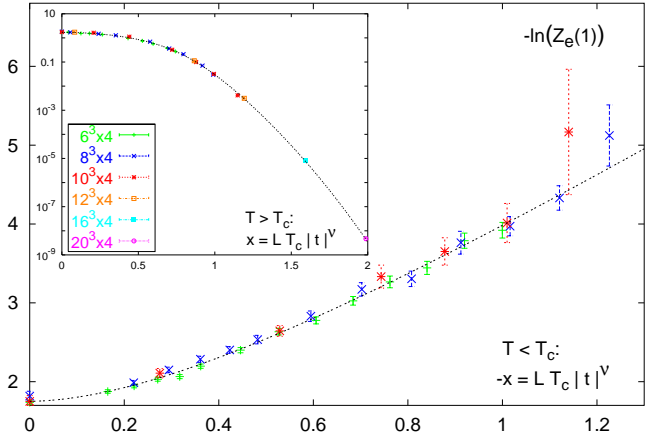,width=1.05\linewidth}}
\vskip -.5cm
\caption{Free energy of one unit of electric flux from
Eq.~\ref{Ze1}  as a function of $x$ 
in the confined phase, and above $T_c$ (insert).} 
 \label{E1} 
\end{figure}

Above $T_c$, the dual string tension is
(with $\tilde\sigma_0^{\scriptscriptstyle
(1)} =  b_+^{\scriptscriptstyle (1)}$),  
\begin{equation}
   \tilde\sigma(T) = \tilde\sigma_0^{\scriptscriptstyle (1)} 
          \, T_c^2 \, |t|^{2\nu} =
   R/\xi_+^2(t) \; , \label{Rratio}
\end{equation}
where the universal ratio $R \simeq  0.104 $ \cite{Kle92,Has97}   
is known from the 3d-Ising model. There, $R = \xi_-^2 \sigma_I$ relates 
the \linebreak correlation length and the interface tension $\sigma_I$ 
for $T\!<T_c$. Here, 
Eq.~(\ref{Rratio}) determines the screening length for the Polyakov loops 
above $T_c$, $\xi_+(t) = \sqrt{R/ \tilde\sigma(T)}$.
In addition, the universality hypothesis relates the ratio 
of the correlation lengths for the Polyakov loops in $SU(2)$ to 
that of their dual analogue, the correlation lengths of the spins in the 
3d-Ising model, as measured in Ref.~\cite{Has97},
\begin{equation}
\label{uar}
 \xi_-^{SU(2)}/\xi_+^{SU(2)}  \,  \stackrel{!}{=} \, 
    \xi_+^{\mbox{\tiny Ising}}/\xi_-^{\mbox{\tiny Ising}} \,  \simeq \, 1.96
    \; .
\end{equation}    
Together with (\ref{Rratio}) this relates the string tension amplitude
below to its dual counter part above $T_c$, as follows. 
From the linear part of the electric-flux free energy,
\begin{eqnarray}
\label{st}
\big\langle P(\vec x) P^\dagger(\vec x +L\vec e_i) \big\rangle 
      &\to &  e^{-\sigma(T) L/T} =
      e^{-L/\xi_-(t)}   , \;\; T<T_c  \nonumber \\
\Rightarrow \;  -\ln(Z_e(1)) &\to &  -x/(\xi_-^{(0)} T_c) \; , \;   
            -x = L T_c  |t|^\nu     ,   
\end{eqnarray} 
for large $L$ (or $(-x)$ large). Thus, from
Eqs.~(\ref{Rratio}), (\ref{uar}),
\[  \frac{\sigma(T)}{T} = \frac{1}{\xi_-(t)} =   T_c  \, | t |^\nu 
                       \sqrt{\tilde\sigma_0^{(1)}/R_+}
       \; , \;  R_+ = \frac{\xi_-^2}{\xi_+^2}\,  R \simeq 0.4
\]
A value of about $\tilde\sigma_0^{\scriptscriptstyle (1)} \simeq 5.36$ then
implies for the string-tension amplitude $1/(\xi_-^{\scriptscriptstyle
(0)}T_c)  \simeq  3.66$. 
This value was used in the fit for the slope of the linear part of 
$-\ln(Z_e(1))$ shown in Fig.~\ref{E1}. 
Fitting the slope to the data yields the consistent value
$1/(\xi_-^{\scriptscriptstyle (0)}T_c) = 3.58 \pm 0.5$. 
This is the linear potential between static charges
without Coulomb part.

The insert of Fig.~\ref{E1} shows the quality of the finite size scaling 
from the more accurate data for the free energy of one electric flux  
in the high temperature phase.

\begin{figure}[t]
\vspace*{-.7cm}
\rightline{\epsfig{file=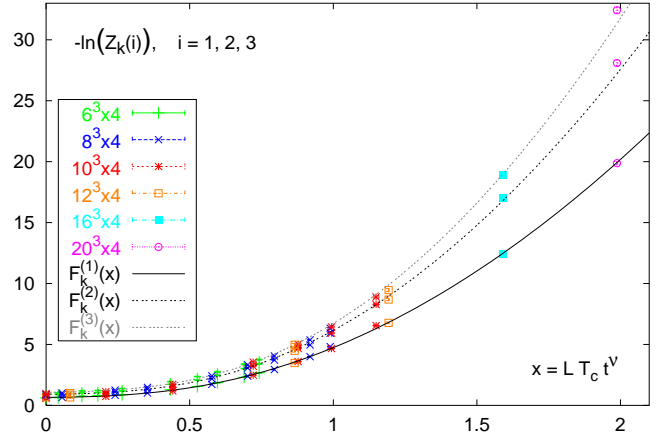,width=1.05\linewidth}}
\vskip -.5cm
\caption{Free energy of 1, 2 and 3 't~Hooft loops at $T > T_c$.}   
\label{tHallhot} 
\end{figure}

Similar results are obtained from Eqs.~(\ref{Ze2}), (\ref{Ze3}) also
for 2 and 3 orthogonal electric fluxes which we verify to be
suppressed more strongly in the confined phase, $Z_e(2)/Z_e(1),
Z_e(3)/Z_e(1) \to 0$ for $L\to\infty$.
Then, inverting Eqs. (\ref{Ze1})-(\ref{Ze3}) one therefore deduces
\begin{eqnarray}
       \hspace*{.2cm} 
    && -\ln (Z_k(1)) \to -\ln(1-2Z_e(1)) \sim Z_e(1) \; , \;\; T<T_c 
           \nonumber\\
     &&  \Rightarrow   
                \quad  1/{(\xi_-^{(0)}T_c)} = b_-^{(1)}  = 3.87\pm 0.5 \; 
\label{Zk-app}
\end{eqnarray}
for the string tension, which is again consistent with the 
value $\simeq 3.66$ implied by the universal amplitude ratio. 
Due to similar relations for
$Z_k(2)$ and $Z_k(3)$, we expect $b_-^{\scriptscriptstyle (1)} =
b_-^{\scriptscriptstyle (2)} = 
b_-^{\scriptscriptstyle (3)}$ in our fits (\ref{f+-}) for 2 and 3 spatial
't~Hooft loops below $T_c$, which is verified well within our ${\cal O}(10\%)$
accuracy 
on these amplitudes. Above $T_c$ on the other hand, we expect for the dual string tension amplitudes
\begin{equation}
 \tilde\sigma_0^{(1)} : \tilde\sigma_0^{(2)} : \tilde\sigma_0^{(3)} 
      \; \sim \;  1 : \sqrt{2}:\sqrt{3}  \; ,   
\end{equation}
according to the effective area of diagonal loops.
In Fig.~\ref{tHallhot} such a square-root
behavior is successfully enforced on the fits (\ref{f+-}) of the spatial
't~Hooft loops above $T_c$. 
We obtain the same ratios, with less accuracy,  
for the slopes of the electric-flux free energies below $T_c$, 
as expected for diagonal fluxes with string formation.

To summarize, we have shown that, below $T_c$, the free 
energy of electric fluxes diverges linearly with
the length $L$ of the system. Because spatial twists
share their qualitative low-temperature behavior with 
the temporal ones considered here, the free energy of the magnetic fluxes
must vanish. 
This is the magnetic Higgs phase with electric confinement
of $SU(2)$ Yang-Mills theory.

At criticality all free energies rapidly approach their finite 
$L\to\infty$ limits indicative of massless excitations. 
We obtain, {\it e.g.}, $Z_k(1) = 0.54(1)$ for $T=T_c$, which agrees with 
the corresponding ratio in the 3d-Ising model \cite{Has97}.

Above $T_c$, 
the free energy of electric charges vanishes in the
thermodynamic limit.  
The dual area law prevents magnetic charges from propagating in spatial
directions. 

The transition is well described by exponents and amplitude ratios of the
3d-Ising class, see also \cite{Pep01}.

We are grateful to P.~van Baal, M.~Garc\'{\i}a P\'erez, 
F.~Gliozzi, C.~Korthals-Altes, A.~Kovner, J.~Pawlow\-ski, M. Pepe and
H.~Reinhardt for discussions. L.v.S. is indebted to the CERN
Theory Division for his visiting appointment during which this project was
initiated.

\end{document}